# Consensus on Transaction Commit

Jim Gray and Leslie Lamport

Microsoft Research

1 January 2004
revised 19 April 2004

**Abstract**

The distributed transaction commit problem requires reaching agreement on whether a transaction is committed or aborted. The classic Two-Phase Commit protocol blocks if the coordinator fails. Fault-tolerant consensus algorithms also reach agreement, but do not block whenever any majority of the processes are working. The Paxos Commit algorithm runs a Paxos consensus algorithm on the commit/abort decision of each participant to obtain a transaction commit protocol that uses $2F + 1$ coordinators and makes progress if at least $F + 1$ of them are working. Paxos Commit has the same stable-storage write delay, and can be implemented to have the same message delay in the fault-free case, as Two-Phase Commit, but it uses more messages. The classic Two-Phase Commit algorithm is obtained as the special $F = 0$ case of the Paxos Commit algorithm.

# Contents

# 1 Introduction

A distributed transaction consists of a number of operations, performed at multiple sites, terminated by a request to commit or abort the transaction. The sites then use a transaction commit protocol to decide whether the transaction is committed or aborted. The transaction can be committed only if all sites are willing to commit it. Achieving this all-or-nothing atomicity property in a distributed system is not trivial. The requirements for transaction commit are stated precisely in Section 2.

The classic transaction commit protocol is Two-Phase Commit [7], described in Section 3. It uses a single coordinator to reach agreement. The failure of that coordinator can cause the protocol to block, with no process knowing the outcome, until the coordinator is repaired. In Section 4, we use the Paxos consensus algorithm [10] to obtain a transaction commit protocol that uses multiple coordinators; it makes progress if a majority of the coordinators are working. Section 5 compares Two-Phase Commit and Paxos Commit. We show that Two-Phase Commit is a degenerate case of the Paxos Commit algorithm with a single coordinator, guaranteeing progress only if that coordinator is working.

Our computation model assumes that algorithms are executed by a collection of processes that communicate using messages. Each process executes at a node in a network. Different processes may execute on the same node. Our cost model counts inter-node message delays; we assume that messages between processes on the same node have negligible delay. Our failure model assumes that nodes, and hence their processes, can fail; messages can be lost or duplicated, but not (undetectably) corrupted. Any process executing at a failed node simply stops performing actions; it does not perform incorrect actions and does not forget its state. Implementing this model of process failure requires writing information to stable storage, which can be an expensive operation. We will see that the delays incurred by writes to stable storage are the same in Two-Phase Commit and Paxos Commit.

In general, there are two kinds of correctness properties that an algorithm must satisfy: safety and liveness. Intuitively, a safety property describes what is allowed to happen, and a liveness property describes what must happen [1].

Our algorithms are asynchronous in the sense that their safety properties do not depend on timely execution by processes or on bounded message delay. Progress, however, may depend on how quickly processes respond and messages are delivered.

We define a nonfaulty node to be one whose processes respond to mes-sages within some known time limit. A network of nodes is nonfaulty iff all its nodes are nonfaulty and messages sent between processes running on those nodes are delivered within some time limit. We will not attempt to formalize the progress properties of our algorithms or the definition of nonfaulty.

The main body of this paper informally describes transaction commit and our two protocols. The Appendix contains formal TLA$^+$ [12] specifications of their safety properties—that is, specifications omitting assumptions and requirements involving progress or real-time constraints. We expect that only the most committed readers will look at those specifications.

## 2  Transaction Commit

In a distributed system, a transaction is performed by a collection of processes called resource managers (RMs), each executing on a different node. The transaction ends when one of the resource managers issues a request either to commit or to abort the transaction. For the transaction to be committed, each participating RM must be willing to commit it. Otherwise, the transaction must be aborted. Prior to the commit request, any RM may spontaneously decide to abort its part of the transaction. The fundamental requirement is that all RMs must agree on whether the transaction is committed or aborted.[1]

To participate, an RM must first join the transaction. For now, we assume a fixed set of participating RMs determined in advance. Section 6.2 discusses how RMs join the transaction.

We abstract the requirements of a transaction commit protocol as follows. (The requirements are summarized in the state-transition diagram of Figure 1.) We assume a set of RM processes, each beginning in a *working* state. The goal of the protocol is for the RMs all to reach a *committed* state or all to reach an *aborted* state. Two safety requirements of the protocol are:

**Stability** Once an RM has entered the *committed* or *aborted* state, it remains in that state forever.

**Consistency** It is impossible for one RM to be in the *committed* state and another to be in the *aborted* state.

---

[1] In some descriptions of transaction commit, there is a client process that ends the transaction and must also learn if it is committed. We consider such a client to be one of the RMs.

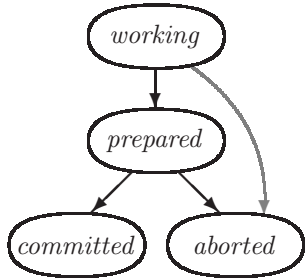

Figure 1: The state-transition diagram for a resource manager. It begins in the *working* state, in which it may decide that it wants to abort or commit. It aborts by simply entering the *aborted* state. If it decides to commit, it enters the *prepared* state. From this state, it can commit only if all other resource managers also decided to commit.

These two properties imply that, once an RM enters the *committed* state, no other RM can enter the *aborted* state, and vice versa.

Each RM also has a *prepared* state. We require that

- An RM can enter the *committed* state only after all RMs have been in the *prepared* state.

These requirements imply that the transaction can commit, meaning that all RMs reach the *committed* state, only by the following sequence of events:

- All the RMs enter the *prepared* state, in any order.

- All the RMs enter the *committed* state, in any order.

The protocol allows the following event that prevents the transaction from committing:

- Any RM in the *working* state can enter the *aborted* state.

The stability and consistency conditions imply that this spontaneous abort event cannot occur if some RM has entered the *committed* state. In practice, an RM will abort when it learns that a failure has occurred that prevents the transaction from committing. However, our abstract representation of the problem permits an RM spontaneously to enter the *aborted* state.

The goal of the algorithm is for all RMs to reach the *committed* or *aborted* state, but this cannot be achieved in a non-trivial way if RMs can fail or become isolated through communication failure. (A trivial solution is one in which all RMs always abort.) Moreover, the classic theorem of Fischer, Lynch, and Paterson [6] implies that a deterministic, purely asynchronous algorithm cannot satisfy the stability and consistency conditions and still guarantee progress in the presence of even a single fault. We therefore require progress only if timeliness hypotheses are satisfied. Our two liveness requirements for a transaction commit protocol are:

**Non-Triviality** If the entire network is nonfaulty throughout the execution of the protocol, then (a) if all RMs reach the *prepared* state, then all RMs reach the *committed* state, and (b) if some RM reaches the *aborted* state, then all RMs reach the *aborted* state.

**Non-Blocking** If, at any time, a sufficiently large network of nodes is nonfaulty for long enough, then every RM executed on those nodes will reach either the *committed* or *aborted* state.

A precise statement of these two conditions would require a precise definition of what it means for a network of nodes to be nonfaulty. The meaning of "long enough" in the Non-Blocking condition depends on the response times of nonfaulty processes and communication networks. We will not attempt to formulate the Non-Triviality and Non-Blocking conditions precisely.

We can more precisely specify a transaction commit protocol by specifying its set of legal behaviors, where a behavior is a sequence of system states. We specify the safety properties with an initial predicate and a next-state relation that describes all possible steps (state transitions). The initial predicate asserts that all RMs are in the *working* state. To define the next-state relation, we first define two state predicates:

**canCommit** True iff all RMs are in the *prepared* or *committed* state.

**notCommitted** True iff no RM is in the *committed* state.

The next-state relation asserts that each step consists of one of the following two actions performed by a single RM:

**Prepare** The RM can change from the *working* state to the *prepared* state.

**Decide** If the RM is in the *prepared* state and *canCommit* is true, then it can transition to the *committed* state; and if the RM is in either the *working* or *prepared* state and *notCommitted* is true, then it can transition to the *aborted* state.

## 3 Two-Phase Commit

### 3.1 The Protocol

The Two-Phase Commit protocol is an implementation of transaction commit that uses a *transaction manager* (TM) process to coordinate the decision-making procedure. The RMs have the same states in this protocol as in the specification of transaction commit. The TM has the following states: *init* (its initial state), *preparing*, *committed*, and *aborted*.

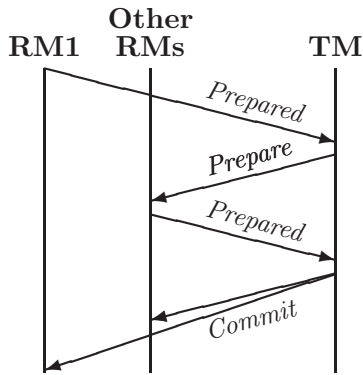

Figure 2: The message flow for Two-Phase Commit in the normal failure-free case, where RM1 is the first RM to enter the *prepared* state.

The Two-Phase Commit protocol starts when an RM enters the *prepared* state and sends a *Prepared* message to the TM. Upon receipt of the *Prepared* message, the TM enters the *preparing* state and sends a *Prepare* message to every other RM. Upon receipt of the *Prepare* message, an RM that is still in the *working* state can enter the *prepared* state and send a *Prepared* message to the TM. When it has received a *Prepared* message from all RMs, the TM can enter the *committed* state and send *Commit* messages to all the other processes. The RMs can enter the *committed* state upon receipt of the *Commit* message from the TM. The message flow for the Two-Phase Commit protocol is shown in Figure 2.

Figure 2 shows one distinguished RM spontaneously preparing. In fact, any RM can spontaneously go from the *working* to *prepared* state and send a *prepared* message at any time. The TM's *prepare* message can be viewed as an optional suggestion that now would be a good time to do so. Other events, including real-time deadlines, might cause working RMs to prepare. This observation is the basis for variants of the Two-Phase Commit protocol that use fewer messages.

An RM can spontaneously enter the *aborted* state if it is in the *working* state; and the TM can spontaneously enter the *aborted* state unless it is in the *committed* state. When the TM aborts, it sends an *abort* message to all RMs. Upon receipt of such a message, an RM enters the *aborted* state. In an implementation, spontaneous aborting can be triggered by a timeout.[2]

Two-Phase Commit is described in many texts [2]; we will not bother to prove its correctness. The protocol is specified formally in Section A.2 of the Appendix, along with a theorem asserting that it implements the specification of transaction commit. This theorem has been checked by the TLC model checker for large enough configurations (numbers of RMs) so it is unlikely to be incorrect.

---

[2]In practice, an RM may notify the TM when it spontaneously aborts; we ignore this optimization.

## 3.2 The Cost of Two-Phase Commit

The important efficiency measure for a transaction commit protocol is the cost of the normal case, in which the transaction is committed. Let $N$ be the number of RMs. The Two-Phase Commit protocol sends the following sequence of messages in the normal case:

- The initiating RM enters the prepared state and sends a *Prepared* message to the TM. (1 message)

- The TM sends a *Prepare* message to every other RM. ($N-1$ messages)

- Each other RM sends a *Prepared* message to the TM. ($N-1$ messages)

- The TM sends a *Commit* message to every RM. ($N$ messages)

Thus, in the normal case, the RMs learn that the transaction has been committed after four message delays. A total of $3N - 1$ messages are sent. It is typical for the TM to be on the same node as the initiating RM. In that case, two of the messages are intra-node and can be discounted, leaving $3N - 3$ messages.

As discussed in Section 3.1, we can eliminate the TM's *Prepare* messages, reducing the message complexity to $2N$. But in practice, this requires either extra message delays or some real-time assumptions.

In addition to the message delays, the two-phase commit protocol incurs the delays associated with three writes to stable storage: the write by the first RM to prepare, the writes by the remaining RMs when they prepare, and the write by the TM when it makes the commit decision. This can be reduced to two writes by having all RMs prepare concurrently.

## 3.3 The Problem with Two-Phase Commit

In a transaction commit protocol, if one or more RMs fail, the transaction is usually aborted. For example, in the Two-Phase Commit protocol, if the TM does not receive a *Prepared* message from some RM soon enough after sending the *Prepare* message, then it will abort the transaction by sending *Abort* messages to the other RMs. However, the failure of the TM can cause the protocol to block until the TM is repaired. In particular, if the TM fails after every RM has sent a *Prepared* message, then the other RMs have no way of knowing whether the TM committed or aborted the transaction.

A non-blocking commit protocol is one in which the failure of a single process does not prevent the other processes from deciding if the transaction is committed or aborted. Several such protocols have been proposed, and a few have been implemented. They have usually attempted to "fix" the Two-Phase Commit protocol by choosing another TM if the first TM fails. However, we know of none that provides a complete algorithm proven to satisfy a clearly stated correctness condition. For example, the discussion of non-blocking commit in the classic text of Bernstein, Hadzilacos, and Goodman [2] fails to explain what a process should do if it receives messages from two different processes, both claiming to be the current TM. Guaranteeing that this situation cannot arise is a problem that is as difficult as implementing a transaction commit protocol.

## 4 Paxos Commit

### 4.1 The Paxos Consensus Algorithm

The distributed computing community has studied the more general problem of *consensus*, which requires that a collection of processes agree on some value. Many solutions to this problem have been proposed, under various failure assumptions [5, 16]. These algorithms have precise fault models and rigorous proofs of correctness.

In the consensus problem, a collection of processes called *acceptors* cooperate to choose a value. Each acceptor runs on a different node. The basic safety requirement is that only a single value be chosen. To rule out trivial solutions, there is an additional requirement that the chosen value must be one proposed by a client. The liveness requirement asserts that, if a large enough subnetwork of the acceptors' nodes is nonfaulty for a long enough time, then some value is eventually chosen. It can be shown that, without strict synchrony assumptions, $2F + 1$ acceptors are needed to achieve consensus despite the failure of any $F$ of them.

The Paxos algorithm [4, 10, 11, 13] is a popular asynchronous consensus algorithm. It assumes some method of choosing a coordinator process, called the *leader*. Clients send proposed values to the leader. In normal operation, there is a unique leader. A new leader is selected only when the current one fails. However, a unique leader is needed only to ensure progress; safety is guaranteed even if there is no leader or there are multiple leaders.

The Paxos algorithm satisfies the liveness requirement for consensus if the leader-selection algorithm ensures that a unique nonfaulty leader is chosen whenever a large enough subnetwork of the acceptors' nodes is nonfaulty for a long enough time.

The algorithm uses a set of ballot numbers, which we take to be nonnegative integers. Each ballot number "belongs to" a unique possible leader.

There is a predetermined choice of initial leader. In the normal, failure-free case, when the leader receives a proposed value, it sends a phase 2a message to all acceptors containing this value and ballot number 0. (The missing phase 1 is explained below.) Each acceptor receives this message, and replies with a phase 2b message for ballot number 0. When the leader receives these phase 2b messages from a majority of acceptors, it sends a phase 3 message announcing that the value is chosen.

The initial leader may not succeed in getting a value chosen in ballot 0, perhaps because it fails. In that case, one or more additional ballots are executed. The following is the general algorithm for executing a ballot; it is performed whenever a new leader is selected.

**Phase 1a** The leader chooses a ballot number *bal* that belongs to it and that it thinks is larger than any ballot number for which phase 1 has been performed. The leader sends a phase 1a message for ballot number *bal* to every acceptor.

**Phase 1b** When an acceptor receives the phase 1a message for ballot number *bal*, if it has not already performed any action for a ballot numbered *bal* or higher, it responds with a phase 1b message containing its current state, which consists of

- The largest ballot number for which it received a phase 1a message, and
- The phase 2b message with the highest ballot number it has sent, if any.

**Phase 2a** When the leader has received a phase 1b message for ballot number *bal* from a majority of the acceptors, it can learn one of two possibilities:

**Free** The algorithm has not yet chosen a value.

**Forced** The algorithm might already have chosen a particular value $v$.

In the free case, the leader can try to get any value accepted; it usually picks the first value proposed by a client. In the forced case, it must try to get the value $v$ chosen. It tries to get a value chosen by sending a phase 2a message with that value and with ballot number *bal* to every acceptor.

**Phase 2b** When an acceptor receives a phase 2a message for a value $v$ and ballot number *bal*, if it has not already received a phase 1a or 2a message for a larger ballot number, it *accepts* that message and sends a phase 2b message for $v$ and *bal* to the leader.

**Phase 3** When the leader has received phase 2b messages for value $v$ and ballot *bal* from a majority of the acceptors, it knows that the value $v$ has been chosen and communicates that fact to all interested processes with a phase 3 message.

In the normal fault-free case, described above, the algorithm starts with the initial leader having already performed phase 1 for ballot number 0. Since there are no ballot numbers less than 0, acceptors cannot have done anything for such ballot numbers and hence have nothing to report in phase 1 for ballot number 0.

The Paxos consensus algorithm can be optimized in two independent ways. We can reduce the number of messages in the normal fault-free case by having the leader send phase 2a messages only to a majority of the acceptors. The leader will know that value $v$ is chosen if it receives phase 2b messages from that majority of acceptors. It can send phase 2a messages to additional acceptors if it does not receive enough phase 2b messages. The second optimization is to eliminate the message delay of phase 3, at the cost of extra messages, by having acceptors send their phase 2b messages directly to all processes that need to know the chosen value. Like the leader, those processes learn the chosen value when they receive phase 2b messages from a majority of the acceptors.

A complete specification of the Paxos algorithm must also describe how a leader interprets phase 1b messages to determine if, and for what value, it is in the forced state. The reader can find such a description in the literature [4, 10, 11, 13]. It also appears in the definition of the *Phase2a* action in the formal specification of our Paxos Commit algorithm that appears in Section A.3 of the Appendix.

The Paxos algorithm guarantees that at most one value is chosen despite any non-malicious failure of any part of the system—that is, as long as processes do not make errors in executing the algorithm and the communication network does not undetectably corrupt messages. It guarantees progress if a unique leader is selected and if the network of nodes executing that leader and some majority of acceptors is nonfaulty for a long enough period of time. A precise statement and proof of this progress condition has been given by De Prisco, Lampson, and Lynch [4].

In practice, it is not difficult to construct an algorithm that, except during rare periods of network instability, selects a suitable unique leader among a majority of nonfaulty acceptors. Transient failure of the leader-selection algorithm is harmless, violating neither safety nor eventual progress.

## 4.2 The Paxos Commit Algorithm

In the Two-Phase Commit protocol, the TM decides whether to abort or commit, records that decision in stable storage, and informs the RMs of its decision. We could make that fault-tolerant by simply using a consensus algorithm to choose the *committed/aborted* decision. (An early instance of this approach was by Mohan, Strong, and Finkelstein [14].) However, in the normal case, the leader must learn that each RM has prepared before it can try to get the value *committed* chosen. Having the RMs tell the leader that they have prepared requires at least one message delay. Our *Paxos Commit* algorithm eliminates that message delay as follows.

Paxos Commit uses a separate instance of the Paxos consensus algorithm to obtain agreement on the decision each RM makes of whether to prepare or abort—a decision we represent by the values *Prepared* and *Aborted*. So, there is one instance of Paxos for each RM. The transaction is committed iff each RM's instance of the consensus algorithm chooses *Prepared*; otherwise the transaction is aborted.

The same set of $2F + 1$ acceptors and the same leader is used for each instance of Paxos. We assume for now that the RMs know the acceptors in advance. In ordinary Paxos, a ballot 0 phase 2a message can have any value $v$. While the leader usually sends such a message, the Paxos algorithm works just as well if, instead of the leader, some other single process sends that message. In Paxos Commit, each RM announces its prepare/abort decision by sending, in its instance of Paxos, a ballot 0 phase 2a message with the value *Prepared* or *Aborted*.

Execution of Paxos Commit normally starts when some RM decides to prepare and sends a *BeginCommit* message to the leader. The leader then sends a *Prepare* message to all the other RMs. If an RM decides that it wants to prepare, it sends a phase 2a message with value *Prepared* and ballot number 0 in its instance of the Paxos algorithm. Otherwise, it sends a phase 2a message with the value *Aborted* and ballot number 0. For each instance, an acceptor sends its phase 2b message to the leader. The leader knows the outcome of this instance if it receives $F + 1$ phase 2b messages for ballot number 0, whereupon it can send its phase 3 message announcing the outcome to the RMs. (As observed in Section 4.1 above, phase 3 can be eliminated by having the acceptors send their phase 2b messages directly to the RMs.) The transaction is committed iff every RM's instance of the Paxos algorithm chooses *Prepared*; otherwise the transaction is aborted.

For efficiency, an acceptor can bundle its phase 2b messages for all instances of the Paxos algorithm into a single physical message. The leader can distill its phase 3 messages for all instances into a single *Commit* or *Abort* message, depending on whether or not all instances chose the value *Prepared*.

The instances of the Paxos algorithm for one or more RMs may not reach a decision with ballot number 0. In that case, the leader (alerted by a timeout) assumes that each of those RMs has failed and executes phase 1a for a larger ballot number in each of their instances of Paxos. If, in phase 2a, the leader learns that its choice is free (so that instance of Paxos has not yet chosen a value), then it tries to get *Aborted* chosen in phase 2b.

An examination of the Paxos algorithm—in particular, of how the decision is reached in phase 2a—shows that the value *Prepared* can be chosen in the instance for resource manager $rm$ only if $rm$ sends a phase 2a message for ballot number 0 with value *Prepared*. If $rm$ instead sends a phase 2a message for ballot 0 with value *Aborted*, then its instance of the Paxos algorithm can choose only *Aborted*, which implies that the transaction must be aborted. In this case, Paxos Commit can short-circuit and inform all processes that the transaction has aborted. This short-circuiting is possible only for phase 2a messages with ballot number 0. It is possible for an instance of the Paxos algorithm to choose the value *Prepared* even though a leader has sent a phase 2a message (for a ballot number greater than 0) with value *Aborted*.

We briefly sketch an intuitive proof of correctness of Paxos Commit. Recall that, in Section 2, we stated that a non-blocking algorithm should satisfy four properties: Stability, Consistency, Non-Triviality, and Non-Blocking. The algorithm satisfies Stability because once an RM receives a decision from a leader, it never changes its view of what value has been chosen. Consistency holds because each instance of the Paxos algorithm chooses a unique value, so different leaders cannot send different decisions. Non-Triviality holds if the leader waits long enough before performing phase 1a for a new ballot number so that, if there are no failures, then each Paxos instance will finish performing phase 2 for ballot 0. The Non-Blocking property follows from the Paxos progress property, which implies that each instance of Paxos eventually chooses either *Prepared* or *Aborted* if a large enough network of acceptors is nonfaulty. More precisely, the Non-Blocking property holds if Paxos satisfies the liveness requirement for consensus, which is the case if the leader-selection algorithm ensures that a unique nonfaulty leader is chosen whenever a large enough subnetwork of the acceptors' nodes is nonfaulty for a long enough time.

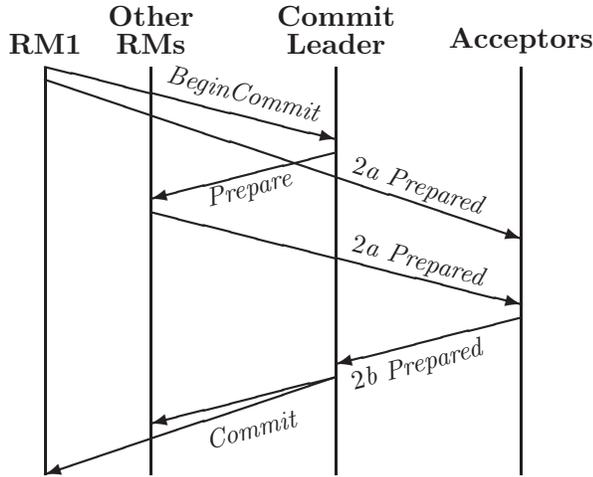

Figure 3: The message flow for Paxos Commit in the normal failure-free case, where RM1 is the first RM to enter the prepared state, and 2a *Prepared* and 2b *Prepared* are the phase 2a and 2b messages of the Paxos consensus algorithm.

The safety part of the algorithm—that is, the algorithm with no progress requirements—is specified formally in Section A.3 of the Appendix, along with a theorem asserting that it implements transaction commit. The correctness of this theorem has been checked by the TLC model checker on configurations that are too small to detect subtle errors, but are probably large enough to find simple "coding" errors. Rigorous proofs of the Paxos algorithm convince us that it harbors no subtle errors, and correctness of the Paxos Commit algorithm seems to be a simple corollary of the correctness of Paxos.

## 4.3 The Cost of Paxos Commit

We now consider the cost of Paxos Commit in the normal case, when the transaction is committed. The sequence of message exchanges is shown in Figure 3.

We again assume that there are $N$ RMs. We consider a system that can tolerate $F$ faults, so there are $2F + 1$ acceptors. However, we assume the optimization in which the leader sends phase 2a messages to $F + 1$ acceptors, and only if one or more of them fail are other acceptors used. In the normal case, the Paxos Commit algorithm uses the following potentially inter-node

messages:

- The first RM to prepare sends a *BeginCommit* message to the leader. (1 message)

- The leader sends a *Prepare* message to every other RM. ($N-1$ messages)

- Each RM sends a ballot 0 phase 2a *Prepared* message for its instance of Paxos to the $F+1$ acceptors. ($N(F+1)$ messages)

- For each RM's instance of Paxos, an acceptor responds to a phase 2a message by sending a phase 2b *Prepared* message to the leader. However, an acceptor can bundle the messages for all those instances into a single message. ($F$ messages, since the leader is an acceptor)

- The leader sends a single *Commit* message to each RM containing a phase 3 *Prepared* message for every instance of Paxos. ($N$ messages)

The RMs therefore learn after five message delays that the transaction has been committed. A total of $(N+1)(F+3)-2$ messages are sent. If the leader is on the same node as one of the acceptors, then one of the phase 2b messages is free, and the first RM's *BeginCommit* message can be accompanied by a phase 2a message, so the total number of messages is $(N+1)(F+3)-4$. If each acceptor is on the same node as an RM, with the first RM being on the same node as the leader, then the messages between the first RM and the leader and an additional $F$ of the phase 2a messages are intra-node and can be discounted, leaving $N(F+3)-3$ messages.

As observed above, we can eliminate phase 3 of Paxos by having each acceptor send its phase 2b messages directly to all the RMs. This allows the RMs to learn the outcome in only four message delays, but a total of $N(2F+3)$ messages are required. Letting the leader be on the same node as an acceptor eliminates one of those messages. If each acceptor is on the same node as an RM, and the leader is on the same node as the first RM, then the initial *BeginCommit* message, $F+1$ of the phase 2a messages, and $F+1$ of the phase 2b messages can be discounted, leaving $(N-1)(2F+3)$ messages.

We have seen so far that Paxos Commit requires five message delays, which can be reduced to four by eliminating phase 3 and having acceptors send extra phase 2b messages. Two of those message delays result from the sending of *Prepare* messages to the RMs. As observed in Section 3.1, these delays can be eliminated by allowing the RMs to prepare spontaneously, leaving just two message delays. This is optimal because implementing transaction commit requires reaching consensus on an RM's decision, and it can be shown that any fault-tolerant consensus algorithm requires at least two message delays to choose a value [3]. The only previous algorithm that achieves the optimal message delay of the optimized version of Paxos Commit is by Guerraoui, Larrea, and Schiper [9].

The RMs perform the same writes to stable storage in Paxos Commit as in Two-Phase Commit. In the Paxos consensus algorithm, an acceptor must record in stable storage its decision to send a phase 2b message before actually sending it. Paxos Commit does this with a single write for all instances of the consensus algorithm. This write corresponds to the TM's write to stable storage before sending a *Commit* message in Two-Phase Commit. Paxos Commit therefore has the same delay caused by writing to stable storage as Two-Phase Commit, and it performs a total of $N + F + 1$ writes.

## 5  Paxos versus Two-Phase Commit

In the Two-Phase Commit protocol, the TM both makes the abort/commit decision and stores that decision in stable storage. Two-Phase Commit is not non-blocking because the stable storage is not accessible if the TM fails. Had we used Paxos simply to obtain consensus on a single decision value, this would have been equivalent to replacing the TM's stable storage by the acceptors' stable storage, and replacing the single TM by a set of possible leaders. Our Paxos Commit algorithm goes further in essentially eliminating the TM's role in making the decision. In Two-Phase Commit, the TM can unilaterally decide to abort. In Paxos Commit, the leader can only make an *abort* decision for an RM that does not decide for itself, which it does by initiating a ballot with number greater than 0 for that RM's instance of Paxos. (The leader must be able to do this to prevent blocking by a failed RM.)

Sections 3.2 and 4.3 describe the cost in messages and writes to stable storage of Two-Phase Commit and Paxos Commit, respectively. Both algorithms have the same three stable write delays (two if all RMs prepare concurrently). The other costs are summarized in Figure 4. The entries for Paxos Commit assume that the leader is on the same node as an acceptor. Faster Paxos Commit is the algorithm optimized to remove phase 3 of the Paxos consensus algorithm. For Two-Phase Commit, co-location means that the initiating RM and the TC are on the same node. For Paxos Com-mit, it means that each acceptor is on the same node as an RM, and that the initiating RM is the on the same node as the leader. In Paxos Commit without co-location, we assume that the leader is an acceptor.

|                        | Two-Phase Commit | Paxos Commit        | Faster Paxos Commit |
|------------------------|:----------------:|:-------------------:|:-------------------:|
| Message Delays         | 4                | 5                   | 4                   |
| Messages               |                  |                     |                     |
|   no co-location   | $3N-1$     | $(N+1)(F+3)-4$      | $N(2F+3)-1$         |
|   with co-location | $3N-3$     | $N(F+3)-3$          | $(N-1)(2F+3)$       |
| Writes to stable storage | $N+1$          | $N+F+1$             | $N+F+1$             |

Figure 4: Corresponding Message Complexity

For the near future, system designers are likely to be satisfied with a commit algorithm that is non-blocking despite at most one failure—the $F = 1$ case. For a transaction with 5 RMs, the Two-Phase Commit then uses 12 messages, regular Paxos Commit uses 17, and Faster Paxos Commit uses 20 (with co-location). For larger values of $N$, the three algorithms use about $3N$, $4N$, and $5N$ messages, respectively (with or without co-location).

Consider now the trivial case of Paxos Commit with $F = 0$, so there is just a single acceptor and the algorithm does not tolerate any acceptor faults. (The algorithm can still tolerate RM faults.) Let the single acceptor and the leader be on the same node. The single phase 2b message of the Paxos consensus algorithm then serves as a phase 3 message, making phase 3 unnecessary. Paxos Commit therefore becomes the same as Faster Paxos Commit. Figure 4 shows that Two-Phase Commit and Paxos Commit use the same number of messages, $3N - 1$ or $3N - 3$, depending on whether or not co-location is assumed. In fact, Two-Phase Commit and Paxos Commit are essentially the same when $F = 0$. The two algorithms are isomorphic under the following correspondence:

| Two-Phase Commit | | Paxos Commit |
|---|---|---|
| TM | ↔ | acceptor/leader |
| *Prepare* message | ↔ | *Prepare* message |
| *Prepared* message | ↔ | phase 2a *Prepared* message |
| *Commit* message | ↔ | *Commit* message |
| *Aborted* message | ↔ | phase 2a *Aborted* message |
| *Abort* message | ↔ | *Abort* message |

The phase 2b/phase 3 *Aborted* message that corresponds to a TM *abort* message is one generated by any instance of the Paxos algorithm, indicating that the transaction is aborted because not all instances chose *Prepared*. The phase 1 and 2 messages that precede it are all sent between the leader and the acceptor, which are on the same node.

The Two-Phase Commit protocol is thus the degenerate case of the Paxos Commit algorithm with a single acceptor.

# 6 Transaction Creation and Registration

So far, we have been considering a single transaction with a fixed set of participating resource managers. In a real system, an application creates a transaction and then invokes RMs that must join the transaction before participating in it. We now discuss how a transaction is created, and how RMs join.

## 6.1 Creation

An application creates a transaction by calling a *transaction service*. Typically, this is done with an API call that returns a *transaction* object with methods such as *join*, *prepared*, and *abort*. The object's private fields identify the processes that execute these method calls. For Two-Phase Commit, the only such process is the TM. In Paxos Commit, the acceptors and the leader take part in the execution.

The API call that creates the transaction object chooses the processes to execute the object's methods and notifies them of the object's creation. The choice of processes can depend on parameters to the API call. In particular, the degree of fault tolerance desired can be a parameter that determines the number of acceptors for Paxos Commit.

In a transaction commit protocol, an RM can receive a message such as *Prepare* that is not a response to a message it sent. In a transaction service API, such a message is implemented as a callback to a method provided by the RM as an argument to *join*.

The basic problems of implementing a transaction commit protocol as a transaction service are the same for the Two-Phase Commit protocol, with its single TM process, as for Paxos Commit, with its multiple participating processes.

## 6.2 Registration

After creating a transaction object, an application invokes one or more RMs. An RM participates in the transaction commit protocol by invoking methods on the transaction object, first invoking *join*. It may eventually invoke *prepared* or *abort*.

The transaction object also has a *close* method, which causes subsequent invocations of *join* to fail. The application can invoke the *close* method when it knows that all RMs needed to perform the transaction have joined. The *close* method might also be used to start the commit process, signaling all RMs that have joined to either prepare or abort. The transaction can be committed only if *close* has been invoked and every RM that successfully joined the transaction has invoked *prepare*.

In Two-Phase Commit, the TM maintains the set of RMs that have joined the transaction. The *join* and *close* methods are invoked by sending a message to the TM. The TM can commit the transaction only if it has received a *prepared* invocation message from every RM that joined.

Paxos Commit also uses a single process, which we call the *registrar*, to maintain the set of RMs that have joined. (This process will usually be located on the same node as the initially chosen leader.) However, Paxos Commit uses an additional instance of the Paxos consensus algorithm to obtain agreement on the set of joined RMs, even if the registrar fails. The registrar participates in this instance of the Paxos consensus algorithm by sending a ballot 0 phase 2a message—just as the RMs do for their instances.

Invoking the *new*, *join*, or *close* method sends a message to the registrar. When the registrar receives a *close* invocation message, it sends a ballot 0 phase 2a message whose value $v$ is the set $S$ of RMs that have joined the transaction. The transaction has committed iff the registrar's instance of Paxos has chosen $S$, and the instance of Paxos for each RM in $S$ has chosen *Prepared*.

As with the RMs, if the registrar's instance of Paxos does not choose a value in ballot number 0 (usually because the registrar has failed), the leader executes phase 1a for a larger ballot number. If the leader learns that it is free to get any value chosen, it chooses a special failure value $\bot$. The transaction has aborted if the registrar's instance of Paxos chooses the value $\bot$. Failure of the registrar can therefore cause the transaction to abort. However, failure of a single RM can also cause the transaction to abort. Fault-tolerance means only that failure of an individual process does not prevent a commit/abort decision from being made.

# 7  Conclusion

Two-Phase Commit is the classical transaction commit protocol. Indeed, it is sometimes thought to be synonymous with transaction commit [15]. Two-Phase Commit is not fault tolerant because it uses a single coordinator whose failure can cause the protocol to block. We have introduced Paxos Commit, a new transaction commit protocol that uses multiple coordinators and makes progress if a majority of them are working. Hence, $2F + 1$ coordinators can make progress even if $F$ of them are faulty. Two-Phase Commit is isomorphic to Paxos Commit with a single coordinator.

In the normal, failure-free case, Paxos Commit requires one more message delay than Two-Phase Commit. Faster Paxos Commit, the version optimized to reduce message delay, has the theoretically minimal message delay for a non-blocking protocol. This low message delay is also achieved by an algorithm of Guerraoui, Larrea, and Schiper [9]. That algorithm is essentially the same as Faster Paxos Commit in the absence of failures. (An optimization analogous to the sending of phase 2a messages only to a majority of acceptors gives it the same message complexity as Faster Paxos Commit.) However, their algorithm is conceptually more complicated than Paxos Commit, using a special procedure for the failure-free case and calling upon a normal consensus algorithm in the event of failure.

With $2F + 1$ coordinators and $N$ resource managers, Paxos Commit requires about $2FN$ more messages than Two-Phase Commit in the normal case. Both algorithms incur the same delay for writing to stable storage. In modern local area networks, messages are cheap, and the cost of writing to stable storage can be much larger than the cost of sending messages. So in many systems, the benefit of a non-blocking protocol should outweigh the additional cost of Paxos Commit.

Paxos Commit implements transaction commit with the Paxos consensus algorithm. Some readers may find this paradoxical, since there are results in the distributed systems theory literature showing that transaction commit is a strictly harder problem than consensus [8]. However, those results are based on a stronger definition of transaction commit in which the transaction is required to commit if all RMs are nonfaulty and choose to prepare—even in the face of unpredictable communication delays. In contrast, our Non-Triviality condition requires the transaction to commit only under the additional assumption that the entire network is nonfaulty—meaning that all messages sent between the nodes are delivered within some known time limit. (Guerraoui, Larrea, and Schiper stated this condition more abstractly in terms of failure detectors.) The stronger definition of transaction commit is not implementable in typical transaction systems, where occasional long communication delays must be tolerated.

# References

[1] Bowen Alpern and Fred B. Schneider. Defining liveness. *Information Processing Letters*, 21(4):181–185, October 1985.

[2] Philip A. Bernstein, V. Hadzilacos, and Nathan Goodman. *Concurrency Control and Recovery in Database Systems*. Addison-Wesley, Reading, Massachusetts, 1987.

[3] Bernadette Charron-Bost and André Schiper. Uniform consensus is harder than consensus (extended abstract). Technical Report DSC/2000/028, École Polytechnique Fédérale de Lausanne, Switzerland, May 2000.

[4] Roberto De Prisco, Butler Lampson, and Nancy Lynch. Revisiting the Paxos algorithm. In Marios Mavronicolas and Philippas Tsigas, editors, *Proceedings of the 11th International Workshop on Distributed Algorithms (WDAG 97)*, volume 1320 of *Lecture Notes in Computer Science*, pages 111–125, Saarbruken, Germany, 1997. Springer-Verlag.

[5] Cynthia Dwork, Nancy Lynch, and Larry Stockmeyer. Consensus in the presence of partial synchrony. *Journal of the ACM*, 35(2):288–323, April 1988.

[6] Michael J. Fischer, Nancy Lynch, and Michael S. Paterson. Impossibility of distributed consensus with one faulty process. *Journal of the ACM*, 32(2):374–382, April 1985.

[7] Jim Gray. Notes on data base operating systems. In R. Bayer, R. M. Graham, and G. Seegmuller, editors, *Operating Systems: An Advanced Course*, volume 60 of *Lecture Notes in Computer Science*, pages 393–481. Springer-Verlag, Berlin, Heidelberg, New York, 1978.

[8] Rachid Guerraoui. Revisiting the relationship between non-blocking atomic commitment and consensus. In Jean-Michel Hélary and Michel Raynal, editors, *Proceedings of the 9th International Workshop on Distributed Algorithms (WDAG95)*, volume 972 of *Lecture Notes in Computer Science*, pages 87–100, Le Mont-Saint-Michel, France, September 1995. Springer-Verlag.

[9] Rachid Guerraoui, Mikel Larrea, and André Schiper. Reducing the cost for non-blocking in atomic commitment. In *Proceedings of the 16th International Conference on Distributed Computing Systems (ICDCS)*, pages 692–697, Hong Kong, May 1996. IEEE Computer Society.

[10] Leslie Lamport. The part-time parliament. *ACM Transactions on Computer Systems*, 16(2):133–169, May 1998.

[11] Leslie Lamport. Paxos made simple. *ACM SIGACT News (Distributed Computing Column)*, 32(4 (Whole Number 121)):18–25, December 2001.

[12] Leslie Lamport. *Specifying Systems*. Addison-Wesley, Boston, 2003. A link to an electronic copy can be found at `http://lamport.org`.

[13] Butler W. Lampson. How to build a highly available system using consensus. In Ozalp Babaoglu and Keith Marzullo, editors, *Distributed Algorithms*, volume 1151 of *Lecture Notes in Computer Science*, pages 1–17, Berlin, 1996. Springer-Verlag.

[14] C. Mohan, R. Strong, and S. Finkelstein. Method for distributed transaction commit and recovery using byzantine agreement within clusters of processors. In *Proceedings of the Second Annual ACM Symposium on Principles of Distributed Computing*, pages 29–43. The Association for Computing Machinery, 1983.

[15] Eric Newcomer. *Understanding Web Services*. Addison-Wesley, Boston, 2002.

[16] Marshall Pease, Robert Shostak, and Leslie Lamport. Reaching agreement in the presence of faults. *Journal of the ACM*, 27(2):228–234, April 1980.

# A  The TLA⁺ Specifications

## A.1  The Specification of a Transaction Commit Protocol

──────────────── MODULE *TCommit* ────────────────

CONSTANT $RM$    The set of participating resource managers
VARIABLE $rmState$    $rmState[rm]$ is the state of resource manager $rm$.

────────────────────────────────────────────

$TCTypeOK \triangleq$
  The type-correctness invariant
  $rmState \in [RM \to \{\text{``working''}, \text{``prepared''}, \text{``committed''}, \text{``aborted''}\}]$

$TCInit \triangleq \ rmState = [rm \in RM \mapsto \text{``working''}]$
  The initial predicate.

$canCommit \triangleq \forall rm \in RM : rmState[rm] \in \{\text{``prepared''}, \text{``committed''}\}$
  True iff all RMs are in the "prepared" or "committed" state.

$notCommitted \triangleq \forall rm \in RM : rmState[rm] \neq \text{``committed''}$
  True iff no resource manager has decided to commit.

────────────────────────────────────────────

We now define the actions that may be performed by the RMs, and then define the complete next-state action of the specification to be the disjunction of the possible RM actions.

$Prepare(rm) \triangleq \ \wedge rmState[rm] = \text{``working''}$
                     $\wedge rmState' = [rmState \text{ EXCEPT } ![rm] = \text{``prepared''}]$

$Decide(rm) \triangleq \ \vee \wedge rmState[rm] = \text{``prepared''}$
                      $\wedge canCommit$
                      $\wedge rmState' = [rmState \text{ EXCEPT } ![rm] = \text{``committed''}]$
                $\vee \wedge rmState[rm] \in \{\text{``working''}, \text{``prepared''}\}$
                      $\wedge notCommitted$
                      $\wedge rmState' = [rmState \text{ EXCEPT } ![rm] = \text{``aborted''}]$

$TCNext \triangleq \exists rm \in RM : Prepare(rm) \vee Decide(rm)$
  The next-state action.

────────────────────────────────────────────

$TCSpec \triangleq \ TCInit \wedge \Box[TCNext]_{rmState}$
  The complete specification of the protocol.

────────────────────────────────────────────

We now assert invariance properties of the specification.

$TCCConsistent \triangleq$

A state predicate asserting that two RMs have not arrived at conflicting decisions.

$$\forall\, rm1,\, rm2 \in RM : \neg \wedge rmState[rm1] = \text{``aborted''} \\ \wedge rmState[rm2] = \text{``committed''}$$

THEOREM $TCSpec \Rightarrow \Box(TCTypeOK \wedge TCCConsistent)$

Asserts that *TCTypeOK* and *TCInvariant* are invariants of the protocol.

## A.2 The Specification of the Two-Phase Commit Protocol

──────── MODULE *TwoPhase* ────────

This specification describes the Two-Phase Commit protocol, in which a transaction manager (TM) coordinates the resource managers (RMs) to implement the Transaction Commit specification of module *TCommit*. In this specification, RMs spontaneously issue *Prepared* messages. We ignore the *Prepare* messages that the TM can send to the RMs.

For simplicity, we also eliminate *Abort* messages sent by an RM when it decides to abort. Such a message would cause the TM to abort the transaction, an event represented here by the TM spontaneously deciding to abort.

This specification describes only the safety properties of the protocol–that is, what is allowed to happen. What must happen would be described by liveness properties, which we do not specify.

CONSTANT $RM$  The set of resource managers

VARIABLES
  $rmState$,    $rmState[rm]$ is the state of resource manager RM.
  $tmState$,    The state of the transaction manager.
  $tmPrepared$, The set of RMs from which the TM has received "*Prepared*" messages.
  $msgs$

In the protocol, processes communicate with one another by sending messages. Since we are specifying only safety, a process is not required to receive a message, so there is no need to model message loss. (There's no difference between a process not being able to receive a message because the message was lost and a process simply ignoring the message.) We therefore represent message passing with a variable *msgs* whose value is the set of all messages that have been sent. Messages are never removed from *msgs*. An action that, in an implementation, would be enabled by the receipt of a certain message is here enabled by the existence of that message in *msgs*. (Receipt of the same message twice is therefore allowed; but in this particular protocol, receiving a message for the second time has no effect.)

$Message \triangleq$

The set of all possible messages. Messages of type "Prepared" are sent from the RM indicated by the message's $rm$ field to the TM. Messages of type "Commit" and "Abort" are broadcast by the TM, to be received by all RMs. The set $msgs$ contains just a single copy of such a message.

$[type : \{\text{"Prepared"}\},\ rm : RM]\ \cup\ [type : \{\text{"Commit"},\ \text{"Abort"}\}]$

$TPTypeOK\ \triangleq$

The type-correctness invariant

$\quad \wedge\ rmState \in [RM \rightarrow \{\text{"working"},\ \text{"prepared"},\ \text{"committed"},\ \text{"aborted"}\}]$
$\quad \wedge\ tmState \in \{\text{"init"},\ \text{"committed"},\ \text{"aborted"}\}$
$\quad \wedge\ tmPrepared \subseteq RM$
$\quad \wedge\ msgs \subseteq Message$

$TPInit\ \triangleq$

The initial predicate.

$\quad \wedge\ rmState = [rm \in RM \mapsto \text{"working"}]$
$\quad \wedge\ tmState = \text{"init"}$
$\quad \wedge\ tmPrepared\ = \{\}$
$\quad \wedge\ msgs = \{\}$

---

We now define the actions that may be performed by the processes, first the TM's actions, then the RMs' actions.

$TMRcvPrepared(rm)\ \triangleq$

The TM receives a "Prepared" message from resource manager $rm$.

$\quad \wedge\ tmState = \text{"init"}$
$\quad \wedge\ [type \mapsto \text{"Prepared"},\ rm \mapsto rm] \in msgs$
$\quad \wedge\ tmPrepared' = tmPrepared \cup \{rm\}$
$\quad \wedge\ \text{UNCHANGED}\ \langle rmState,\ tmState,\ msgs \rangle$

$TMCommit\ \triangleq$

The TM commits the transaction; enabled iff the TM is in its initial state and every RM has sent a "Prepared" message.

$\quad \wedge\ tmState = \text{"init"}$
$\quad \wedge\ tmPrepared = RM$
$\quad \wedge\ tmState' = \text{"committed"}$
$\quad \wedge\ msgs' = msgs \cup \{[type \mapsto \text{"Commit"}]\}$
$\quad \wedge\ \text{UNCHANGED}\ \langle rmState,\ tmPrepared \rangle$

$TMAbort\ \triangleq$

The TM spontaneously aborts the transaction.

$\quad \wedge\ tmState = \text{"init"}$

$\quad \wedge\ tmState' =$ "aborted"
$\quad \wedge\ msgs' = msgs \cup \{[type \mapsto$ "Abort"$]\}$
$\quad \wedge\ $ UNCHANGED $\langle rmState,\ tmPrepared \rangle$

$RMPrepare(rm) \triangleq$

Resource manager $rm$ prepares.

$\quad \wedge\ rmState[rm] =$ "working"
$\quad \wedge\ rmState' = [rmState$ EXCEPT $![rm] =$ "prepared"$]$
$\quad \wedge\ msgs' = msgs \cup \{[type \mapsto$ "Prepared", $rm \mapsto rm]\}$
$\quad \wedge\ $ UNCHANGED $\langle tmState,\ tmPrepared \rangle$

$RMChooseToAbort(rm) \triangleq$

Resource manager $rm$ spontaneously decides to abort. As noted above, $rm$ does not send any message in our simplified spec.

$\quad \wedge\ rmState[rm] =$ "working"
$\quad \wedge\ rmState' = [rmState$ EXCEPT $![rm] =$ "aborted"$]$
$\quad \wedge\ $ UNCHANGED $\langle tmState,\ tmPrepared,\ msgs \rangle$

$RMRcvCommitMsg(rm) \triangleq$

Resource manager $rm$ is told by the TM to commit.

$\quad \wedge\ [type \mapsto$ "Commit"$] \in msgs$
$\quad \wedge\ rmState' = [rmState$ EXCEPT $![rm] =$ "committed"$]$
$\quad \wedge\ $ UNCHANGED $\langle tmState,\ tmPrepared,\ msgs \rangle$

$RMRcvAbortMsg(rm) \triangleq$

Resource manager $rm$ is told by the TM to abort.

$\quad \wedge\ [type \mapsto$ "Abort"$] \in msgs$
$\quad \wedge\ rmState' = [rmState$ EXCEPT $![rm] =$ "aborted"$]$
$\quad \wedge\ $ UNCHANGED $\langle tmState,\ tmPrepared,\ msgs \rangle$

$TPNext \triangleq$
$\quad \vee\ TMCommit \vee TMAbort$
$\quad \vee\ \exists\, rm \in RM :$
$\quad\quad\quad TMRcvPrepared(rm) \vee RMPrepare(rm) \vee RMChooseToAbort(rm)$
$\quad\quad\quad\quad \vee\ RMRcvCommitMsg(rm) \vee RMRcvAbortMsg(rm)$

---

$TPSpec \triangleq TPInit \wedge \Box[TPNext]_{\langle rmState,\ tmState,\ tmPrepared,\ msgs \rangle}$

The complete spec of the Two-Phase Commit protocol.

THEOREM $TPSpec \Rightarrow \Box TPTypeOK$

This theorem asserts that the type-correctness predicate TPTypeOK is an invariant of the specification.

We now assert that the Two-Phase Commit protocol implements the Transaction Commit protocol of module *TCommit*. The following statement defines *TC!TCSpec* to be formula *TSpec* of module *TCommit*. (The TLA$^+$ INSTANCE statement is used to rename the operators defined in module *TCommit* avoids any name conflicts that might exist with operators in the current module.)

$TC \triangleq$ INSTANCE *TCommit*

THEOREM $TPSpec \Rightarrow TC!TCSpec$

This theorem asserts that the specification TPSpec of the Two-Phase Commit protocol implements the specification TCSpec of the Transaction Commit protocol.

The two theorems in this module have been checked with TLC for six RMs, a configuration with 50816 reachable states, in a little over a minute on a 1 GHz PC.

## A.3 The Paxos Commit Algorithm

──────── MODULE *PaxosCommit* ────────

This module specifies the Paxos Commit algorithm. We specify only safety properties, not liveness properties. We simplify the specification in the following ways.

- As in the specification of module *TwoPhase*, and for the same reasons, we let the variable *msgs* be the set of all messages that have ever been sent. If a message is sent to a set of recipients, only one copy of the message appears in *msgs*.

- We do not explicitly model the receipt of messages. If an operation can be performed when a process has received a certain set of messages, then the operation is represented by an action that is enabled when those messages are in the set *msgs* of sent messages. (We are specifying only safety properties, which assert what events can occur, and the operation can occur if the messages that enable it have been sent.)

- We do not model leader selection. We define actions that the current leader may perform, but do not specify who performs them.

As in the specification of Two-Phase commit in module *TwoPhase*, we have RMs spontaneously issue Prepared messages and we ignore *Prepare* messages.

EXTENDS *Integers*

$Maximum(S) \triangleq$

If $S$ is a set of numbers, then this define $Maximum(S)$ to be the maximum of those numbers, or $-1$ if $S$ is empty.

LET $Max[T \in \text{SUBSET } S] \triangleq$
$\quad$ IF $T = \{\}$ THEN $-1$
$\quad\quad\quad$ ELSE LET $n \quad \triangleq$ CHOOSE $n \in T :$ TRUE
$\quad\quad\quad\quad\quad\quad\quad rmax \triangleq Max[T \setminus \{n\}]$

$\quad$ IN $\quad$ IF $n \geq rmax$ THEN $n$ ELSE $rmax$

IN $\quad Max[S]$

CONSTANT $RM$, $\quad\quad$ The set of resource managers.
$\quad\quad\quad\quad Acceptor$, $\quad\quad$ The set of acceptors.
$\quad\quad\quad\quad Majority$, $\quad\quad$ The set of majorities of acceptors
$\quad\quad\quad\quad Ballot$ $\quad\quad\quad$ The set of ballot numbers

ASSUME $\quad$ We assume these properties of the declared constants.
$\quad \wedge\ Ballot \subseteq Nat$
$\quad \wedge\ 0 \in Ballot$
$\quad \wedge\ Majority \subseteq$ SUBSET $Acceptor$
$\quad \wedge\ \forall\ MS1,\ MS2 \in Majority : MS1 \cap MS2 \neq \{\}$

> All we assume about the set *Majority* of majorities is that any two majorities have non-empty intersection.

$Message \triangleq$

> The set of all possible messages. There are messages of type "Commit" and "Abort" to announce the decision, as well as messages for each phase of each instance of *ins* of the Paxos consensus algorithm. The *acc* field indicates the sender of a message from an acceptor to the leader; messages from a leader are broadcast to all acceptors.

$[type : \{\text{``phase1a''}\},\ ins : RM,\ bal : Ballot \setminus \{0\}]$
$\quad\quad \cup$
$[type : \{\text{``phase1b''}\},\ ins : RM,\ mbal : Ballot,\ bal : Ballot \cup \{-1\},$
$\ val : \{\text{``prepared''},\ \text{``aborted''},\ \text{``none''}\},\ acc : Acceptor]$
$\quad\quad \cup$
$[type : \{\text{``phase2a''}\},\ ins : RM,\ bal : Ballot,\ val : \{\text{``prepared''},\ \text{``aborted''}\}]$
$\quad\quad \cup$
$[type : \{\text{``phase2b''}\},\ acc : Acceptor,\ ins : RM,\ bal : Ballot,$
$\ val : \{\text{``prepared''},\ \text{``aborted''}\}]$
$\quad\quad \cup$
$[type : \{\text{``Commit''},\ \text{``Abort''}\}]$

VARIABLES
$\quad rmState$, $\quad rmState[rm]$ is the state of resource manager $rm$.
$\quad aState$, $\quad aState[ins][ac]$ is the state of acceptor $ac$ for instance
$\quad\quad\quad\quad\quad ins$ of the Paxos algorithm
$\quad msgs$ $\quad\quad$ The set of all messages ever sent.

$PCTypeOK \triangleq$

> The type-correctness invariant. Each acceptor maintains the values *mbal*, *bal*, and *val* for each instance of the Paxos consensus algorithm.

$\quad\land rmState \in [RM \to \{\text{"working"}, \text{"prepared"}, \text{"committed"}, \text{"aborted"}\}]$
$\quad\land aState \quad \in [RM \to [Acceptor \to$
$\qquad\qquad\qquad\qquad [mbal : Ballot,$
$\qquad\qquad\qquad\qquad\ bal \quad : Ballot \cup \{-1\},$
$\qquad\qquad\qquad\qquad\ val \quad : \{\text{"prepared"}, \text{"aborted"}, \text{"none"}\}]]]$
$\quad\land msgs \in \text{SUBSET } Message$

$PCInit \triangleq$    The initial predicate.
$\quad\land rmState = [rm \in RM \mapsto \text{"working"}]$
$\quad\land aState \quad = [ins \in RM \mapsto$
$\qquad\qquad\quad [ac \in Acceptor$
$\qquad\qquad\qquad \mapsto [mbal \mapsto 0,\ bal \mapsto -1,\ val \mapsto \text{"none"}]]]$
$\quad\land msgs = \{\}$

## The Actions

$Send(m) \triangleq msgs' = msgs \cup \{m\}$

An action expression that describes the sending of message $m$.

## RM Actions

$RMPrepare(rm) \triangleq$

Resource manager $rm$ prepares by sending a phase 2a message for ballot number 0 with value "prepared".

$\quad\land rmState[rm] = \text{"working"}$
$\quad\land rmState' = [rmState \text{ EXCEPT } ![rm] = \text{"prepared"}]$
$\quad\land Send([type \mapsto \text{"phase2a"},\ ins \mapsto rm,\ bal \mapsto 0,\ val \mapsto \text{"prepared"}])$
$\quad\land \text{UNCHANGED } aState$

$RMChooseToAbort(rm) \triangleq$

Resource manager $rm$ spontaneously decides to abort. It may (but need not) send a phase 2a message for ballot number 0 with value "aborted".

$\quad\land rmState[rm] = \text{"working"}$
$\quad\land rmState' = [rmState \text{ EXCEPT } ![rm] = \text{"aborted"}]$
$\quad\land Send([type \mapsto \text{"phase2a"},\ ins \mapsto rm,\ bal \mapsto 0,\ val \mapsto \text{"aborted"}])$
$\quad\land \text{UNCHANGED } aState$

$RMRcvCommitMsg(rm) \triangleq$

Resource manager $rm$ is told by the leader to commit. When this action is enabled, $rmState[rm]$ must equal either "prepared" or "committed". In the latter case, the action leaves the state unchanged (it is a "stuttering step").

$\quad\land [type \mapsto \text{"Commit"}] \in msgs$
$\quad\land rmState' = [rmState \text{ EXCEPT } ![rm] = \text{"committed"}]$

$\quad\land$ UNCHANGED $\langle aState,\ msgs\rangle$

$RMRcvAbortMsg(rm) \triangleq$

Resource manager *rm* is told by the leader to abort. It could be in any state except "committed".

$\quad\land [type \mapsto \text{"Abort"}] \in msgs$
$\quad\land rmState' = [rmState \text{ EXCEPT } ![rm] = \text{"aborted"}]$
$\quad\land$ UNCHANGED $\langle aState,\ msgs\rangle$

## Leader Actions

The following actions are performed by any process that believes itself to be the current leader. Since leader selection is not assumed to be reliable, multiple processes could simultaneously consider themselves to be the leader.

$Phase1a(bal,\ rm) \triangleq$

If the leader times out without learning that a decision has been reached on resource manager *rm*'s prepare/abort decision, it can perform this action to initiate a new ballot *bal*. (Sending duplicate phase 1a messages is harmless.)

$\quad\land Send([type \mapsto \text{"phase1a"},\ ins \mapsto rm,\ bal \mapsto bal])$
$\quad\land$ UNCHANGED $\langle rmState,\ aState\rangle$

$Phase2a(bal,\ rm) \triangleq$

The action in which a leader sends a phase 2a message with ballot $bal > 0$ in instance *rm*, if it has received phase 1b messages for ballot number *bal* from a majority of acceptors. If the leader received a phase 1b message from some acceptor that had sent a phase 2b message for this instance, then $maxbal \geq 0$ and the value *val* the leader sends is determined by the phase 1b messages. (If $val = $ "prepared", then *rm* must have prepared.) Otherwise, $maxbal = -1$ and the leader sends the value "aborted".

The first conjunct asserts that the action is disabled if any commit leader has already sent a phase 2a message with ballot number *bal*. In practice, this is implemented by having ballot numbers partitioned among potential leaders, and having a leader record in stable storage the largest ballot number for which it sent a phase 2a message.

$\quad\land \lnot \exists\, m \in msgs :\ \land m.type = \text{"phase2a"}$
$\qquad\qquad\qquad\qquad\land m.bal = bal$
$\qquad\qquad\qquad\qquad\land m.ins = rm$
$\quad\land \exists\, MS \in Majority :$
$\qquad\text{LET } mset \triangleq \{m \in msgs :\ \land m.type\ = \text{"phase1b"}$
$\qquad\qquad\qquad\qquad\qquad\qquad\land m.ins\ \ = rm$
$\qquad\qquad\qquad\qquad\qquad\qquad\land m.mbal = bal$
$\qquad\qquad\qquad\qquad\qquad\qquad\land m.acc\ \ \in MS\}$
$\qquad\quad maxbal \triangleq Maximum(\{m.bal : m \in mset\})$
$\qquad\quad val \triangleq \text{ IF } maxbal = -1$
$\qquad\qquad\qquad \text{THEN } \text{"aborted"}$

$$\text{ELSE } (\text{CHOOSE } m \in mset : m.bal = maxbal).val$$

IN  $\quad \wedge \forall\, ac \in MS : \exists\, m \in mset : m.acc = ac$
$\quad\quad \wedge Send([type \mapsto \text{"phase2a"},\ ins \mapsto rm,\ bal \mapsto bal,\ val \mapsto val])$

$\wedge$ UNCHANGED $\langle rmState,\ aState \rangle$

$Decide \triangleq$

> A leader can decide that Paxos Commit has reached a result and send a message announcing the result if it has received the necessary phase 2b messages.

$\wedge$ LET $Decided(rm,\ v) \triangleq$

> True iff instance $rm$ of the Paxos consensus algorithm has chosen the value $v$.

$\quad\quad \exists\, b \in Ballot,\ MS \in Majority :$
$\quad\quad\quad \forall\, ac \in MS : [type \mapsto \text{"phase2b"},\ ins \mapsto rm,$
$\quad\quad\quad\quad\quad\quad\quad\quad\quad bal \mapsto b,\ val \mapsto v,\ acc \mapsto ac] \in msgs$

IN  $\quad \vee\ \wedge \forall\, rm \in RM : Decided(rm,\ \text{"prepared"})$
$\quad\quad\quad \wedge Send([type \mapsto \text{"Commit"}])$
$\quad\quad \vee\ \wedge \exists\, rm \in RM : Decided(rm,\ \text{"aborted"})$
$\quad\quad\quad \wedge Send([type \mapsto \text{"Abort"}])$

$\wedge$ UNCHANGED $\langle rmState,\ aState \rangle$

## Acceptor Actions

$Phase1b(acc) \triangleq$
$\quad \exists\, m \in msgs :$
$\quad\quad \wedge m.type = \text{"phase1a"}$
$\quad\quad \wedge aState[m.ins][acc].mbal < m.bal$
$\quad\quad \wedge aState' = [aState \text{ EXCEPT } ![m.ins][acc].mbal = m.bal]$
$\quad\quad \wedge Send([type\ \mapsto \text{"phase1b"},$
$\quad\quad\quad\quad\quad ins\ \ \mapsto m.ins,$
$\quad\quad\quad\quad\quad mbal \mapsto m.bal,$
$\quad\quad\quad\quad\quad bal\ \ \mapsto aState[m.ins][acc].bal,$
$\quad\quad\quad\quad\quad val\ \ \mapsto aState[m.ins][acc].val,$
$\quad\quad\quad\quad\quad acc\ \ \mapsto acc])$
$\quad\quad \wedge$ UNCHANGED $rmState$

$Phase2b(acc) \triangleq$
$\quad \wedge \exists\, m \in msgs :$
$\quad\quad \wedge m.type = \text{"phase2a"}$
$\quad\quad \wedge aState[m.ins][acc].mbal \leq m.bal$
$\quad\quad \wedge aState' = [aState \text{ EXCEPT } ![m.ins][acc].mbal = m.bal,$
$\quad\quad\quad\quad\quad\quad\quad\quad\quad\quad\ \ ![m.ins][acc].bal\ \ = m.bal,$

$$![m.ins][acc].val = m.val]$$
$$\land Send([type \mapsto \text{``phase2b''}, ins \mapsto m.ins, bal \mapsto m.bal,$$
$$val \mapsto m.val, acc \mapsto acc])$$
$\land$ UNCHANGED $rmState$

$PCNext \triangleq$   The next-state action
$\quad \lor \exists\, rm \in RM : \lor RMPrepare(rm)$
$\qquad\qquad\qquad\quad \lor RMChooseToAbort(rm)$
$\qquad\qquad\qquad\quad \lor RMRcvCommitMsg(rm)$
$\qquad\qquad\qquad\quad \lor RMRcvAbortMsg(rm)$
$\quad \lor \exists\, bal \in Ballot \setminus \{0\}, rm \in RM : Phase1a(bal, rm) \lor Phase2a(bal, rm)$
$\quad \lor Decide$
$\quad \lor \exists\, acc \in Acceptor : Phase1b(acc) \lor Phase2b(acc)$

$PCSpec \triangleq PCInit \land \Box[PCNext]_{\langle rmState,\, aState,\, msgs \rangle}$

The complete spec of the Paxos Commit protocol.

THEOREM $PCSpec \Rightarrow PCTypeOK$

We now assert that the two-phase commit protocol implements the transaction commit protocol of module TCommit. The following statement defines $TC!TCSpec$ to be the formula $TCSpec$ of module $TCommit$. (The TLA$^+$ INSTANCE statement must is used to rename the operators defined in module $TCommit$ to avoid possible name conflicts with operators in the current module having the same name.)

$TC \triangleq$ INSTANCE $TCommit$

THEOREM $PCSpec \Rightarrow TC!TCSpec$